\documentstyle[12pt,a4,epsf]{article}

\def\tr{\mathop{\rm tr}\nolimits}

\makeatletter
\@addtoreset{equation}{section}
\makeatother

\newcommand{\VEV}[1]{\left\langle #1 \right\rangle}

\newcommand{\Mp}{M_P}

\begin{document}
\begin{titlepage}

\begin{flushright}
hep-ph/0110276\\
KUNS-1740\\
\today
\end{flushright}

\vspace{4ex}

\begin{center}
{\large \bf
A Road to the Standard Grand Unified Theory
}

\vspace{6ex}

\renewcommand{\thefootnote}{\alph{footnote}}

Nobuhiro Maekawa\footnote
{e-mail: maekawa@gauge.scphys.kyoto-u.ac.jp
}

\vspace{4ex}
{\it Department of Physics, Kyoto University,\\
     Kyoto 606-8502, Japan}\\
\end{center}

\renewcommand{\thefootnote}{\arabic{footnote}}
\setcounter{footnote}{0}
\vspace{6ex}

%--------------------<<   abstract   >>--------------------
\begin{abstract}

In this talk\footnote{
This talk is based on the works \cite{maekawa,maekawa2,BM}.
}, we propose a GUT scenario in which
doublet-triplet splitting is naturally realized in
$SO(10)$ unification using the Dimopoulos-Wilczek mechanism
\cite{DW}
and the realistic mass matrices of quarks and leptons are
obtained in a simple way. For the neutrino sector, bi-maximal
neutrino mixing angles are realized. 
Moreover, the generic interaction is allowed, namely, all the 
terms which are allowed by the symmetry are included in the 
scenario. Therefore, once we fix the integer number charges
of the anomalous $U(1)_A$ symmetry, which plays an essential
role in the scenario, all the scales, GUT breaking scale,
mass scales of superheavy particles, are determined.
The scenario can be extended into $E_6$ unification, in which
a condition for suppression of flavor changing neutral current
(FCNC) is automatically satisfied.
\end{abstract}

\end{titlepage}

%--------------------<<   section    >>--------------------

\section{Introduction}

In this talk, we propose a scenario of $SO(10)$ grand 
unified theory (GUT) with anomalous $U(1)_A$ gauge symmetry,
 which has the following interesting features\cite{maekawa};
\begin{enumerate}
\item
The doublet-triplet (DT) splitting is realized
using Dimopoulos-Wilczek mechanism
\cite{DW,BarrRaby,Chako,complicate}.
\item
The proton decay via dimension five operator is suppressed.
\item
Realistic quark and lepton mass matrices can be obtained in a simple
way. Especially in neutrino sector, bi-large neutrino mixing is realized.
\item
The symmetry breaking scales are determined by the anomalous $U(1)_A$
charges.
\item
The mass spectrum of the super heavy particles is fixed by the anomalous
$U(1)_A$ charges.
\item The $\mu$ problem is also naturally solved\cite{maekawa2}.
\end{enumerate}

As a consequence of the above features, the fact that the GUT scale is
smaller than the Planck scale
is strongly connected to the improvement of the undesired
GUT relation between the Yukawa couplings 
$y_\mu=y_s$ ($y_e=y_d$ also)
while keeping $y_\tau=y_b$. Moreover, it is remarkable 
that the interaction
is generic, namely, all the interactions, which are allowed 
by the symmetry, are taken into account. Therefore, once we 
fix the field contents with their quantum numbers, all the 
interactions are determined except the coefficients
of order one.
Moreover the above scenario can be extended into $E_6$ 
unification\cite{BM},
in which a suppression condition of FCNC is automatically satisfied.
In $E_6$ unification, the twisting mechanism\cite{Bando} is essential.

There the anomalous $U(1)_A$ gauge symmetry\cite{U(1)},
whose anomaly is cancelled by Green-Schwarz 
mechanism\cite{GS},
 plays an essential role
in the scenario.
%explaining the DT splitting mechanism at the unification
%scale as well as in reproducing Yukawa 
%hierarchies\cite{Ibanez,Ramond,Dreiner}.
%Also bi-large neutrino mixing is naturally
%obtained by choosing ${\bf 10}$ representation with an appropriate
%$U(1)_A$ charge in addition to the three family ${\bf 16}$ representations.
%This anomalous $U(1)_A$  is a powerful tool not only to reproduce
%DT splitting but also to determine the GUT breaking scales.

\section{Relation between VEVs and anomalous $U(1)_A$ charges 
and neutrino masses}
In this section, we explain how the vacua of the Higgs fields are
determined by the anomalous
$U(1)_A$
quantum numbers.\cite{maekawa,BM}

First of all, we show that none of the field with positive anomalous
$U(1)_A$ charge gets nonzero VEV if the Froggatt-Nielsen (FN) mechanism
works well in the vacuum.
 Let the gauge singlet fields be $Z_i^\pm$ ($i=1,2,\cdots n_\pm$)
 with charges $z_i^\pm$ with $z_i^+>0$ and $z_i^-<0$.
From the $F$ flatness conditions of the superpotential
we get $n=n_++n_-$ equations  plus one $D$-flatness condition,
\begin{equation}
 \frac{\delta  W}{\delta Z_i}=0, \qquad D_A=g_A
      \left(\sum_i z_i |Z_i|^2 +\xi^2 \right)=0,
\label{eq:fflat}
\end{equation}
where $\xi^2=\frac{g_s^2\tr Q_A}{192\pi^2} (\equiv \lambda^2 \Mp^2)$.
At a glance, these look to be over determined. However,
 the $F$ flatness
conditions are not independent because the gauge invariance of the
superpotential $W$ leads to a relation
\begin{equation}
\frac{\delta  W}{\delta Z_i}z_iZ_i=0.
\label{constraint}
\end{equation}
Therefore, generically SUSY vacuum with $\VEV{Z_i}\sim \Mp$ exists
(Vacuum a),
because the coefficients of the above conditions are generally
of order 1.
However, if $n_+\leq n_-$, we can take another vacuum (Vacuum b)
with $\VEV{Z_i^+}=0$, which automatically satisfy the $F$-flatness
conditions $\frac{\delta  W}{\delta Z_i^-}=0$. Then $\VEV{Z_i^-}$
are determined by $F$-flatness conditions
$\frac{\delta  W}{\delta Z_i^+}=0$ with a constraint
(\ref{constraint}) and $D$-flatness condition $D_A=0$.
Note that if $\lambda<1$ (i.e., $\xi<1$), 
the VEVs of $Z_i^-$ are less than the
Planck scale, that can lead to Froggatt-Nielsen mechanism.
If we fix the normalization of $U(1)_A$ gauge symmetry so that
the largest value $z_1^-$ in the negative charges $z_i^-$ equals -1
%the maximal value of $z_1^-$ equals 1, 
then the VEV of the field
$Z_1^-$ is determined from $D_A=0$ as $\VEV{Z_1^-}\sim\lambda$,
which breaks $U(1)_A$ gauge symmetry. (The field $Z_1^-$ becomes
the Froggatt-Nielsen field $\Theta$.) On the other hand,
other VEVs are determined by $F$-flatness conditions of $Z_i^+$ as
$\VEV{Z_i^-}\sim \lambda^{-z_i^-}$, which is shown below.
Since $\VEV{Z_i^+}=0$, it is sufficient to examine the terms linear
in $Z_i^+$ in the superpotential in order to determine 
$\VEV{Z_i^-}$. Therefore, in general
the superpotential can be written
\begin{eqnarray}
W&=&\sum_i^{n_+}W_{Z_i^+},\\
W_{Z_i^+}&=& \lambda^{z_i^+}Z_i^+(\sum_j^{n_-}\lambda^{z_j^-}Z_j^-
+\sum_{j,k}^{n_-}\lambda^{z_j^-+z_k^-}Z_j^-Z_k^-+\cdots )\\
&=&\sum_i^{n_+} \tilde Z_i^+(\sum_j^{n_-}\tilde Z_j^-
+\sum_{j,k}^{n_-}\tilde Z_j^-\tilde Z_k^-+\cdots ),
\end{eqnarray}
where $\tilde Z_i\equiv \lambda^{z_i}Z_i$.
The $F$-flatness conditions of $Z_i^+$ fields require
\begin{equation}
\lambda^{z_i^+}(1+\sum_j\tilde Z_j^-+\cdots)=0,
\end{equation}
which generally lead to solutions $\tilde Z_j\sim O(1)$
if these $F$-flatness conditions determine the VEVs.
Thus the F-flatness condition demands,
\begin{equation}
   \VEV{ Z_j} \sim O(\lambda ^{-z_j}).
\label{eq:}
\end{equation}
Here we have examined the VEVs of singlets fields, but generally
the gauge invariant operator $O$ with negative charge $o$ has
non-vanishing VEV $\VEV{O}\sim \lambda^{-o}$ if the $F$-flatness
conditions determine the VEV.
Note that when $n_+=n_-$, all the VEVs $\VEV{Z_i^-}$ can be determined by
the $F$-flatness conditions $\frac{\delta W}{\delta Z_i^+}=0$.
It means that there is no flat direction, namely no massless field.
On the other hand, when $n_+<n_-$, then there must be some 
massless fields related with the flat direction.

If the vacuum a is selected, the anomalous $U(1)_A$ gauge symmetry
is broken at the Planck scale and the FN mechanism does not work.
Therefore, we cannot know the existence of the $U(1)_A$ gauge symmetry
from the low energy physics. On the other hand, if the vacuum b is
selected, the FN mechanism works well and we can understand
the signature of the $U(1)_A$ gauge symmetry from the low energy
physics. Therefore, it is natural to assume that the vacuum b is
selected in our scenario, in which the $U(1)_A$ gauge symmetry
plays an important role for the FN mechanism. Namely, the VEVs of
the fields $Z_i^+$ vanish, that guarantee that the SUSY zero mechanism
works well.

If an adjoint field $A({\bf 45})$ has a VEV by the $F$-flatness 
condition, the scale of the VEV is determined as 
$\VEV{A}\sim \lambda^{-a}$ because $A^2$ can be gauge invariant.
Moreover, in addition to the adjoint field $A$, we have to introduce
spinor Higgs $C{\bf 16})$ and $\bar C({\bf \overline {16}})$ to break
$SO(10)$ into the standard gauge group. The VEV $\VEV{\bar CC}$ are 
determined by the anomalous $U(1)_A$ charges $c+\bar c$ as
$\VEV{\bar CC}=\lambda^{-(c+\bar c)}$. This leads to
\begin{equation}
\VEV{C}=\VEV{\bar C}=\lambda^{-\frac{1}{2}(c+\bar c)}
\end{equation}
because of $D$-flatness condition of $SO(10)$ gauge theory. 
Note that the scale of the VEVs are also determined by the anomalous 
$U(1)_A$ charges, though the relation is different from the naive
expectation $\VEV{C}=\lambda^{-c}$. This is because the $D$-flatness
condition plays a critical role to determine the VEVs.
Note that the power is half integer. This fact plays an important role
to obtain bi-large mixing angles in neutrino sector, which will be 
discussed lator.

\section{Doublet-triplet splitting with anomalous $U(1)_A$ gauge 
symmetry}
In this section, we show that DT splitting is naturally realized
in $SO(10)$ GUT with anomalous $U(1)_A$ gauge symmetry.

The minimal Higgs content to break $SO(10)$ into $SU(3)_C\times U(1)_{EM}$
is one adjoint Higgs $A({\bf 45})$, a pair of spinor fields
$C({\bf 16})$ and $\bar C({\bf \overline{16}})$ and usual Higgs $H({\bf 10})$.
All of them must have negative anomalous $U(1)_A$ charges because they
have non-vanishing VEVs. On the other hand, we have to introduce the same 
number of the fields with positive anomalous $U(1)_A$ charges in order to
make all fields massive\footnote{
Strictly speaking, some component fields are absorbed by the Higgs mechanism,
so we do not have to introduce the same number of the fields with positive
charges. However, it is not the case in $SO(10)$ unification.}.
The content of the Higgs sector with $SO(10)\times U(1)_A$ gauge 
symmetry 
is given in Table I, where the symbols $\pm$ denote a $Z_2$ parity quantum 
numbers.

%\vspace{0.5cm}

\begin{center}
Table I. The lowercase letters represent the anomalous $U(1)_A$ charges.
\begin{eqnarray}
{\bf 45}&:& A(a=-2,-), \quad A^\prime (a^\prime=6,-)  \nonumber \\
{\bf 16}&:& C(c=-4,+), \quad C^\prime (c^\prime=4,-) \nonumber \\
{\bf \overline{16}}&:& \bar C(\bar c=-1,+), \quad 
                  \bar C^\prime (\bar c^\prime=7,-) \nonumber \\
{\bf 10}&:& H(h=-6,+), \quad H^\prime (h^\prime=8,-) \nonumber \\
{\bf 1}&:&  Z(z=-3,-), \quad \bar Z(\bar z=-3,-), \quad S(s=5,+) \nonumber
\end{eqnarray}
\end{center}
Here we have listed typical values of the anomalous
$U(1)_A$ charges. 
Among these fields, $A$, $C$, $\bar C$, $Z$ and $\bar Z$ are expected to
obtain non-vanishing
VEVs
around the GUT scale. As discussed in the previous section, the fields with 
positive $U(1)_A$ charges have vanishing VEVs.
It is surprising that the DT splitting 
mechanism is naturally embedded into the above minimal model in a sense.

Since the fields with non-vanishing VEVs have negative charges, only the 
$F$-flatness conditions of fields with positive charge must be counted
for determination of their VEVs. 
Moreover, we have only to take account of
the terms in the superpotential which contain only one field with 
positive 
charge. This is because the terms with more positive charge fields 
do not 
contribute to the $F$-flatness conditions, since the positive fields 
are
assumed to have zero VEV.
Therefore, in general, the superpotential required by determination 
of the
VEVs can be written as
\begin{equation}
W=W_{H^\prime}+ W_{A^\prime} + W_S + W_{C^\prime}+W_{\bar C^\prime}.
\end{equation}
Here $W_X$ denotes the terms linear in the $X$ field, which has 
positive anomalous $U(1)_A$ charge. Note, however, that terms 
including two
fields with positive charge like 
$\lambda^{2h^\prime}H^\prime H^\prime$
give contributions to the mass terms but not to the VEVs.

We now discuss the determination of the VEVs.
If $-3a\leq a^\prime < -5a$,
the superpotential $W_{A^\prime}$ is in general
written as
\begin{equation}
W_{A^\prime}=\lambda^{a^\prime+a}\alpha A^\prime A+\lambda^{a^\prime+3a}(
\beta(A^\prime A)_{\bf 1}(A^2)_{\bf 1}
+\gamma(A^\prime A)_{\bf 54}(A^2)_{\bf 54}),
\end{equation}
where the suffixes {\bf 1} and {\bf 54} indicate the representation 
of the composite
operators under the $SO(10)$ gauge symmetry, and $\alpha$, $\beta$ and 
$\gamma$ are parameters of order 1. Here we assume 
$a+a^\prime+c+\bar c<0$
to forbid the term $\bar C A^\prime A C$, which destabilizes the 
DW form of the VEV $\VEV{A}$. 
If we take 
$\VEV{A}=i\tau_2\times {\rm diag}(x_1,x_2,x_3,x_4,x_5)$, the $F$-flatness
of the $A^\prime$ field requires
$x_i(\alpha\lambda^{-2a}+2(\beta-\gamma)(\sum_j x_j^2)+\gamma x_i^2)=0$, 
which gives only two solutions $x_i^2=0$, 
$\frac{\alpha}{(2N-1)\gamma-2N\beta}\lambda^{-2a}$. 
Here $N=1-5$ is the number of $x_i \neq 0$ solutions.
The DW form is obtained when $N=3$.
Note that the higher terms $A^\prime A^{2L+1}$ $(L>1)$ are forbidden by
the SUSY zero mechanism. If they are allowed, 
the number of possible VEVs other than the DW form
becomes larger, and thus it becomes less natural to obtain the DW form. 
This is a
critical point of this mechanism, and the anomalous $U(1)_A$ gauge 
symmetry
plays an essential role to forbid the undesired terms.
It is also interesting that the scale of the VEV is automatically
determined by the anomalous $U(1)_A$ charge of $A$, as noted in the 
previous section.

Next we discuss the $F$-flatness condition of $S$, which determines
the scale of the VEV $\VEV{\bar C C}$. 
$W_S$, which is linear in the $S$ field, is given by
\begin{equation}
W_S=\lambda^{s+c+\bar c}S\left((\bar CC)+\lambda^{-(c+\bar c)}
+\sum_k\lambda^{-(c+\bar c)+2ka}A^{2k}\right)
\end{equation}
if 
$s\geq -(c+\bar c)$.
Then the $F$-flatness condition of $S$ implies $\VEV{\bar CC}\sim 
\lambda^{-(c+\bar c)}$, and the $D$-flatness condition requires 
$|\VEV{C}|=|\VEV{\bar C}|\sim \lambda^{-(c+\bar c)/2}$.
The scale of the VEV is determined only by the charges of 
$C$ and $\bar C$ again.
If we take $c+\bar c=-6$, then we obtain the VEVs of the fields 
$\bar C$ and $\bar C$
as 
$\lambda^3$, which differ from the expected values $\lambda^{-c}$ 
and
$\lambda^{-\bar c}$ if $c\neq \bar c$.

Finally, we discuss the $F$-flatness of $C^\prime$ and 
$\bar C^\prime$,
which realizes the alignment of the VEVs $\VEV{C}$ and $\VEV{\bar C}$
and imparts masses on the PNG fields. 
This simple mechanism
was proposed by Barr and Raby
\cite{BarrRaby}.
We can easily assign anomalous $U(1)_A$ charges which allow the 
following superpotential:
\begin{eqnarray}
W_{C^\prime}&=&
       \bar C(\lambda^{\bar c^\prime +c+a}A
       +\lambda^{\bar c^\prime +c+\bar z}\bar Z)C^\prime, \\
W_{\bar C^\prime}&=&
       \bar C^\prime(\lambda^{\bar c^\prime +c+a} A
       +\lambda^{\bar c^\prime +c+z}Z)C.
\end{eqnarray}
The $F$-flatness conditions $F_{C^\prime}=F_{\bar C^\prime}=0$ give
$(\lambda^{a-z} A+Z)C=\bar C(\lambda^{a-\bar z} A+\bar Z)=0$. 
Recall that the VEV of $A$ is 
proportional to the $B-L$ generator $Q_{B-L}$ as 
$\VEV{A}=\frac{3}{2}vQ_{B-L}$.
Also $C$, ${\bf 16}$, is decomposed into 
$({\bf 3},{\bf 2},{\bf 1})_{1/3}$, 
$({\bf \bar 3},{\bf 1},{\bf 2})_{-1/3}$, 
$({\bf 1},{\bf 2},{\bf 1})_{-1}$ and $({\bf 1},{\bf 1},{\bf 2})_{1}$ 
under
$SU(3)_C\times SU(2)_L\times SU(2)_R\times U(1)_{B-L}$.
Since $\VEV{\bar CC}\neq 0$, 
not all components in the spinor $C$ vanish. 
Then $Z$ is fixed to be $Z\sim -\frac{3}{2}\lambda v Q_{B-L}^0$, 
where $Q_{B-L}^0$ is
the $B-L$ charge of the component field in $C$, which has non-vanishing VEV. 
It is interesting
that no other component fields can have non-vanishing VEVs 
because of the $F$-flatness
conditions. If  the $({\bf 1},{\bf 1},{\bf 2})_{\bf 1}$ field obtains
a non-zero VEV (therefore, $\VEV{Z}\sim -\frac{3}{2}\lambda v$), then the 
gauge group 
$SU(3)_C\times SU(2)_L\times SU(2)_R\times U(1)_{B-L}$ is broken to the 
standard gauge group. Once the direction of the VEV $\VEV{C}$ is 
determined, the VEV $\VEV{\bar C}$ must have the same direction 
because of the $D$-flatness
condition. Therefore, $\VEV{\bar Z}\sim -\frac{3}{2}\lambda v$.
Thus, all VEVs have now been fixed.

We do not discuss the detail of the mass spectrum here. 
But all fields acquire the mass term except one pair of doublet
Higgs fields\cite{maekawa}.
We will discuss only the mass matrix of Higgs $H$.
Considering the additional mass term
$\lambda^{2h^\prime} H^\prime H^\prime$, we write the mass matrix of 
the Higgs fields $H$ and $H^\prime$, which are decomposed from {\bf 5} 
and 
${\bf \bar 5}$ of $SU(5)$, as
\begin{equation}
({\bf 5}_H, {\bf 5}_{H^\prime})
\left(\begin{array}{cc} 0 & \lambda^{h+h^\prime +a}\VEV{A} \\
                     \lambda^{h+h^\prime +a}\VEV{A} & \lambda^{2h^\prime}
      \end{array}\right)
\left(\begin{array}{c} {\bf \bar 5}_H \\ {\bf \bar 5}_{H^\prime}
\end{array}\right).
\end{equation}
The colored Higgs obtain their masses of order 
$\lambda^{h+h^\prime+a}\VEV{A}\sim \lambda^{h+h^\prime}$.
Since in general $\lambda^{h+h^\prime}>\lambda^{2h^\prime}$,
the proton decay is naturally suppressed. The effective colored
Higgs mass is estimated as
$(\lambda^{h+h^\prime})^2/\lambda^{2h^\prime}=\lambda^{2h}$, 
which is larger than the Planck scale, because
$h<0$.
One pair of the doublet Higgs is massless,
while another pair of doublet Higgs acquires a mass of order 
$\lambda^{2h^\prime}$. 
The DW mechanism works well, although we have to examine the effect of the 
rather
light additional Higgs.

There are several terms which must be forbidden for the stability of the 
DW mechanism. For example, $H^2$, $HZH^\prime$ and $H\bar Z H^\prime$
induce a large mass of the doublet Higgs, 
and the term $\bar CA^\prime A C$ would destabilize the DW form of 
$\VEV{A}$.
We can easily forbid these terms using the SUSY zero mechanism.
For example, if we choose
$h<0$, then $H^2$ is forbidden, and if we choose $\bar c+c+a+a^\prime<0$, 
then
$\bar CA^\prime A C$ is forbidden. (It is interesting that the negative 
$U(1)_A$ charge $h$, which is required for the DT splitting, enhances
the left-handed neutrino masses, as discussed in section 2.) 
Once these dangerous terms are forbidden
by the SUSY zero mechanism, higher-dimensional terms which also become
dangerous;
for example, 
$\bar CA^\prime A^3 C$ and $\bar CA^\prime C\bar CA C$ are automatically
forbidden, since only gauge invariant operators with negative charge
can have non-vanishing VEVs. This is also an attractive point of our 
scenario. 

In this section, we have proposed an natural DT splitting mechanism 
in which the anomalous $U(1)_A$ gauge symmetry plays a critical role, 
and the VEVs and mass spectrum are automatically determined by the 
anomalous $U(1)_A$ charges. 
In the next section, we examine a model with this DT 
splitting 
mechanism, which gives realistic mass matrices of quarks and leptons.

\section{Quark and Lepton sector}
In this section, we examine the simplest model to demonstrate 
how to determine
everything from the anomalous $U(1)_A$ charges.

In addition to the Higgs sector in Table.I, we introduce only three 
${\bf 16}$ representations $\Psi_i$
with anomalous $U(1)_A$ charges $(\psi_1=n+3,\psi_2=n+2,\psi_3=n)$ 
and one ${\bf 10}$ field
$T$ with
charge $t$ as the matter contents. These matter fields are assigned
odd R-parity, while those of the Higgs sector are assigned even 
R-parity.
Such an assignment of R-parity guarantees that the argument regarding
VEVs 
in the previous section does not change if these matter fields have 
vanishing VEVs.
We can give an argument to determine the allowed region of the 
anomalous $U(1)_A$ charges 
to obtain
desired terms while forbidding dangerous terms. 
Though this is a straightforward
argument, we do not give it here. Instead, 
we give a set of anomalous $U(1)_A$
charges with which all conditions are satisfied and a novel neutrino
mass matrix is obtained:
$n=3, t=4, h=-6, h^\prime=8, c=-4, \bar c=-1, c^\prime=4, 
\bar c^\prime=7, s=5$.
Then the mass term of ${\bf 5}$ and ${\bf \bar 5}$ of $SU(5)$
is written as
\begin{equation}
{\bf 5}_T ( \lambda^6 \VEV{C}, \lambda^5 \VEV{C}, \lambda^3\VEV{C}, 
\lambda^8)
\left(
\begin{array}{c}  {\bf \bar 5}_{\Psi1} \\ {\bf \bar 5}_{\Psi2} \\ 
{\bf \bar 5}_{\Psi3}
 \\ {\bf \bar 5_T}
\end{array}
\right).
\end{equation}
Since $\VEV{\bar C}=\VEV{C}\sim \lambda^{5/2}$, because $c+\bar c=-5$, 
the massive mode ${\bf \bar 5}_M$, the partner of $5_T$, is given by
\begin{equation}
{\bf \bar 5}_M \sim {\bf \bar 5}_{\Psi3}+\lambda^{5/2}{\bf \bar 5}_T.
\end{equation} 
Therefore the three massless modes 
$({\bf \bar 5}_1, {\bf \bar 5}_2, {\bf \bar 5}_3) $ are
written $({\bf \bar 5}_{\Psi1}, 
{\bf \bar 5}_T+ \lambda^{\frac{5}{2}}{\bf \bar 5}_{\Psi3},
{\bf \bar 5}_{\Psi2})$. The Dirac mass matrices for quarks and leptons 
can be 
obtained from the interaction
\begin{equation}
\lambda^{\psi_i+\psi_j+h}\Psi_i\Psi_jH.
\label{Yukawa}
\end{equation}
The mass matrices for the up quark sector and the down quark sector are
\begin{equation}
M_u=\left(
\begin{array}{ccc}
\lambda^6 & \lambda^5 & \lambda^3 \\
\lambda^5 & \lambda^4 & \lambda^2 \\
\lambda^3 & \lambda^2 & 1    
\end{array}
\right)\VEV{H_u},\quad
M_d=\lambda^2\left(
\begin{array}{ccc}
\lambda^4 & \lambda^{7/2} & \lambda^3 \\
\lambda^3 & \lambda^{5/2} & \lambda^2 \\
\lambda^1 & \lambda^{1/2} & 1    
\end{array}
\right)\VEV{H_d}.
\end{equation}
Note that the Yukawa couplings for 
${\bf \bar 5}_2\sim {\bf \bar 5}_T+\lambda^{5/2}{\bf \bar 5}_{\Psi 3}$
are obtained only through the Yukawa couplings for the component 
${\bf \bar 5}_{\Psi 3}$,
because we have no Yukawa couplings for $T$.
We can estimate the CKM matrix from these quark matrices as
\begin{equation}
U_{CKM}=
\left(
\begin{array}{ccc}
1 & \lambda &  \lambda^3 \\
\lambda & 1 & \lambda^2 \\
\lambda^3 & \lambda^2 & 1
\end{array}
\right),
\label{CKM}
\end{equation}
which is consistent with the experimental value if we choose 
$\lambda\sim 0.2$\footnote{
Strictly speaking, if the Yukawa coupling originated only from the interaction
\ref{Yukawa}, the mixing concerning to the first generation becomes smaller than
the expected values because of a cancellation. In order to get the expected
value of CKM matrix as in \ref{CKM}, non-renormalizable terms, for example, 
$\Psi_i\Psi_jH\bar CC$ must be taken into account.}.
Since the ratio of the Yukawa couplings of top and bottom quarks is 
$\lambda^2$,
a small value of $\tan \beta\equiv \VEV{H_u}/\VEV{H_d} \sim O(1)$ is 
predicted by these mass matrices.
The Yukawa matrix for the charged lepton sector is the same as the transpose 
of $M_d$ at this stage, except for an overall factor $\eta$ induced by the 
renormalization group effect. 
The mass matrix for the Dirac mass of neutrinos is given by
\begin{equation}
M_D=\lambda^2\left(
\begin{array}{ccc}
\lambda^4 & \lambda^3 & \lambda \\
\lambda^{7/2} & \lambda^{5/2} & \lambda^{1/2} \\
\lambda^3   & \lambda^2         & 1 
\end{array}
\right)\VEV{H_u}\eta.
\end{equation}

The right-handed neutrino masses come from the interaction
\begin{equation}
\lambda^{\psi_i+\psi_j+2\bar c}\Psi_i\Psi_j\bar C\bar C
\end{equation}
as
\begin{equation}
M_R=\lambda^{\psi_i+\psi_j+2\bar c}\VEV{\bar C}^2=\lambda^9\left(
\begin{array}{ccc}
\lambda^6 & \lambda^5 & \lambda^3 \\
\lambda^5 & \lambda^4 & \lambda^2 \\
\lambda^3   & \lambda^2         & 1
\end{array}
\right).
\end{equation}
Therefore we can estimate the neutrino mass matrix:
\begin{equation}
M_\nu=M_DM_R^{-1}M_D^T=\lambda^{-5}\left(
\begin{array}{ccc}
\lambda^2 & \lambda^{3/2} & \lambda \\
\lambda^{3/2} & \lambda & \lambda^{1/2} \\
\lambda   & \lambda^{1/2}         & 1 
\end{array}
\right)\VEV{H_u}^2\eta^2.
\end{equation}
Note that the overall factor $\lambda^{-5}$ has
negative power, which
can be induced by the effects discussed in sections 2 and 3.
From these mass matrices in the lepton sector the MNS
matrix is obtained as
\begin{equation}
U_{MNS}=
\left(
\begin{array}{ccc}
1 & \lambda^{1/2} &  \lambda \\
\lambda^{1/2} & 1 & \lambda^{1/2} \\
\lambda & \lambda^{1/2} & 1
\end{array}
\right).
\end{equation}
This gives bi-maximal mixing angles for the neutrino sector,
because $\lambda^{1/2}\sim 0.5$.
We then obtain the prediction 
$m_{\nu_\mu}/m_{\nu_\tau}\sim \lambda$, which is consistent with
the experimental data
\cite{SK,SNO}: 
$1.6\times 10^{-3} ({\rm eV})^2\leq \Delta m_{\rm atm}^2\leq 4
\times 10^{-3}
({\rm eV})^2$
and $2\times 10^{-5} ({\rm eV})^2\leq \Delta m_{\rm solar}^2\leq 1
\times 10^{-4}
({\rm eV})^2$.
The relation $V_{e3}\sim \lambda$ is also an interesting 
prediction from this 
matrix, though CHOOZ gives a restrictive upper limit $V_{e3}\leq 0.15$
\cite{CHOOZ}.
The neutrino mass is given by
$m_{\nu_\tau}\sim \lambda^{-5}\VEV{H_u}^2\eta^2/\Mp\sim 
m_{\nu_\mu}/\lambda
\sim m_{\nu_e}/\lambda^2$. If we take $\VEV{H_u}\eta=100$ GeV, 
$\Mp\sim 10^{18}$ GeV and $\lambda=0.2$, then we get 
$m_{\nu_\tau}\sim 3\times 10^{-2}$ eV, $m_{\nu_\mu}\sim 6\times 
10^{-3}$ eV
and $m_{\nu_e}\sim 1\times 10^{-3}$ eV. 
It is surprising that such a rough approximation gives
values in good agreement with
the experimental values from the atmospheric neutrino and large 
mixing angle (LMA) MSW solution of the solar neutrino problem. 
This LMA solution for the solar neutrino problem
gives the best fit to the present experimental data
\cite{Valle}.

In addition to Eq.~(\ref{Yukawa}), the interactions
\begin{equation}
\lambda^{\psi_i+\psi_j+2a+h}\Psi_iA^2\Psi_jH
\end{equation}
also contribute to the Yukawa couplings. Here $A$ is squared because
it has odd parity.
Since $A$ is proportional to the generator of $B-L$,
the contribution to the lepton Yukawa coupling is nine times larger
than that to quark Yukawa coupling, which can change the unrealistic
prediction $m_\mu=m_s$ at the GUT scale. 
Since the prediction $m_s/m_b\sim \lambda^{5/2}$ at the GUT scale is 
consistent with experiment, 
the enhancement factor $2\sim 3$ of $m_\mu$ can improve the situation.
Note that the additional terms contribute mainly in the lepton
sector.
If we set $a=-2$,
the additional matrices are
\begin{eqnarray}
\frac{\Delta M_u}{\VEV{H_u}}&=&
\frac{v^2}{4}\left(
\begin{array}{ccc}
\lambda^2 & \lambda & 0 \\
\lambda & 1 & 0 \\
0 & 0 & 0
\end{array}
\right)
,\quad
\frac{\Delta M_d}{\VEV{H_d}}=
\frac{v^2}{4}\left(
\begin{array}{ccc}
\lambda^2 & 0 & \lambda \\
\lambda & 0 & 1 \\
0 & 0 & 0 
\end{array}
\right)
,
\\
\frac{\Delta M_e}{\VEV{H_d}}&=&
\frac{9v^2}{4}\left(
\begin{array}{ccc}
\lambda^2 & \lambda & 0 \\
0 & 0 & 0 \\
\lambda & 1 & 0 
\end{array}
\right).
\end{eqnarray}
It is interesting that this modification essentially changes the 
eigenvalues of
only the first and second generation. Therefore it is natural to expect 
that a realistic mass pattern can be obtained by this modification.
This is one of the largest motivations to choose $a=-2$.
Note that this charge assignment also determines the scale 
$\VEV{A}\sim \lambda^2$.
It is suggestive that the fact that the $SO(10)$ breaking scale is 
slightly
smaller than the Planck scale is correlated with the discrepancy between
the naive prediction of the ratio $m_\mu/m_s$ from the unification and 
the experimental value\footnote{
Such an argument has been done also in \cite{IKNY}.}. 
It is also interesting that the SUSY zero mechanism plays an essential 
role again. 
When $z, \bar z \geq -4$, the terms 
$\lambda^{\psi_i+\psi_j+a+z+h}Z\Psi_iA\Psi_jH+
\lambda^{\psi_i+\psi_j+2z+h}Z^2\Psi_i\Psi_jH
$ also contribute to the fermion mass matrices, though only to the first
generation.

Proton decay mediated by the colored Higgs is strongly suppressed
in this model. As mentioned in the previous section, the effective 
mass of
the colored Higgs is of order $\lambda^{2h}\sim \lambda^{-12}$, which is 
much larger than the Planck scale. 
Proton decay is also induced by the non-renormalizable
term
\begin{equation} 
\lambda^{\psi_i+\psi_j+\psi_k+\psi_l}\Psi_i\Psi_j\Psi_k\Psi_l,
\end{equation}
which is also strongly suppressed.

%\section{Coupling Unification}
Since the spectrum of the superheavy particles is fixed by anomalous
$U(1)_A$ charges, we can check whether the three gauge couplings are 
unified or not. 
This is a severe constraint to select a realistic model. There is an
example in which the three gauge couplings meet at a scale.
%But at least the coupling unification is possible. 
If we take the anomalous $U(1)_A$ 
charges as 
$\psi_1=5,\psi_2=4,\psi_2=2,t=3,a=-1,a'=3,h=-4,h'=5,c=-3,\bar c=0,c'=2,
\bar c'=5$, then the three gauge couplings are unified at the 
unification scale $\Lambda_G\sim \lambda \Lambda$. Here we have to take
the cutoff scale $\Lambda\sim 2\times 10^{16}$ GeV. Therefore,
in this model, the proton decay due to the dimension 6 operator
may be seen in near future.

\section{A natural solution for the $\mu$ problem}
In our scenario, SUSY zero mechanism forbids the SUSY Higgs mass term
$\mu HH$. However, once SUSY is broken, the Higgs mass $\mu$ must be
induced. The induced mass must be proportional to the SUSY breaking scale.

We now examine a solution for the $\mu$ problem in a simple example
\cite{maekawa2}.
The essential point of this mechanism is that the VEV shift of a heavy
singlet field by SUSY breaking. 
%In the literature
%\cite{hall}, it is shown that the SUSY breaking terms produce the VEV shift
%of heavy particles of order the SUSY breaking scale
%in the context of super gravity scenario.
%If there is a heavy singlet which has vanishing VEV in SUSY limit and
%couples to the 
%Higgs field, then shifting the VEV solves the $\mu$ problem. The argument
%is essentially the same as in Ref.\cite{hempfling}, but they requires
%R-symmetry, which is not a symmetry in our scenario \cite{maekawa}.
%Below we show that such a situation is easily obtained in our scenario,
%namely, R-symmetry is not an essential ingredient of the mechanism.
%Before examining the detail, we figure out the essence of the mechanism.
We introduce the superpotential $W=\lambda^sS+\lambda^{s+z}SZ$, where 
$S$ and $Z$ are singlet fields with positive anomalous $U(1)_A$ charge $s$ 
and with negative charge $z$, respectively ($s+z\geq 0$). 
Note that the single term of $Z$
is not allowed by SUSY zero mechanism, while usual symmetry cannot forbid
this term. This is an essential point of this mechanism. The SUSY vacuum is 
at
$\VEV{S}=0$ and $\VEV{Z}=\lambda^{-z}$. After SUSY is broken, these 
VEVs are modified. To determine the VEV shift of $S$, which we would 
like to know because the singlet $S$ with positive charge can couple to 
the Higgs field with negative charge, the most important SUSY breaking 
term is the tadpole term of $S$, namely $\lambda^s \Mp^2A S$. Here $A$ 
is a SUSY breaking parameter of order of the weak scale. By this tadpole 
term, the VEV of $S$ appears as $\VEV{S}=\lambda^{-s-2z}A$. If we have 
$\lambda^{s+2h} SH^2$, the SUSY Higgs mass is obtained as 
$\mu=\lambda^{2h-2z}m_{SB}$, which is proportional to the SUSY breaking 
parameter $m_{SB}$ and the proportional coefficient can be
of order 1 if $h\sim z$. Note that the $F$-term of
$S$ is calculated as $F_S\sim \lambda^{-s-2z}m_{SB}^2$.
 The Higgs mixing term $B\mu$ can be obtained from the SUSY term
$\lambda^{s+2h}SH^2$ and the
SUSY breaking term $\lambda^{s+2h}A_{SH^2}SH^2$ as
$\lambda^{s+2h}F_S\sim \lambda^{2h-2z}m_{SB}^2$ and 
$\lambda^{2h-2z}A^2\sim \mu A$, respectively.
Therefore the relation $B\sim m_{SB}$ is naturally obtained
\footnote{
If doublet-triplet splitting is realized by fine-tuning or some
accidental cancellation, the Higgs mixing
$B\mu$ can become intermediated scale $m_{SB} M_X$ as discussed in
Ref.\cite{kawamura}, where $M_X$ is the GUT scale.
However, once the 
doublet-triplet splitting is naturally solved  as in 
Ref.\cite{maekawa}, such a problem disappears }.
This is a solution for the $\mu$ problem. Note that the condition 
$h\sim z$ can be satisfied because both fields $H$ and $Z$ have
 negative charges. 

\section{SUSY breaking and FCNC}
We discuss SUSY breaking in this section. Since we should assign the
anomalous $U(1)_A$ charges dependent on the flavor to produce the
hierarchy of Yukawa couplings, generically the
non-degenerate scalar fermion masses are induced through the anomalous
$U(1)_A$ $D$-term.\footnote{
The large SUSY breaking scale can avoid the flavor changing
neutral current (FCNC) problem, but in our scenario
it does not work because the anomalous $U(1)_A$ charge of the Higgs
$H$ is
inevitably negative to forbid the Higgs mass term in tree level.
}
Various experiments on the FCNC processes give
strong constraints
to 
the off-diagonal terms $\Delta$ in the sfermion mass matrices
on the basis on which the flavor changing terms appear only in the 
non-diagonality of the sfermion propagators as in Ref.\cite{masiero}.
The sfermion propagators can be expanded in terms of 
$\delta=\Delta/\tilde m^2$ where $\tilde m$ is an average sfermion mass.
As long as $\Delta$ is sufficiently smaller than $\tilde m^2$, 
it is enough to take the first term of this expansion and, then, 
the experimental
information concerning FCNC and CP violating phenomena is translated into
upper bounds on these $(\delta_{ij}^F)_{XY}$'s, where
$F=U,D,N,E$, the chirality index $X,Y=L,R$ 
and the generation index $i,j=1,2,3$.
For example, 
the experimental value of $K^0-\bar K^0$ mixing gives
\begin{eqnarray}
&&\sqrt{|{\rm Re} (\delta_{12}^D)_{LL}(\delta_{12}^D)_{RR}|}
\leq 2.8\times 10^{-3}
\left( \frac{\tilde m_q ({\rm GeV})}{500}\right), \label{LR}\\
 &&|{\rm Re} (\delta_{12}^D)_{LL}|,|{\rm Re} (\delta_{12}^D)_{RR}|
 \leq 4.0\times 10^{-2}
\left( \frac{\tilde m_q ({\rm GeV})}{500}\right), \label{LL}
\end{eqnarray}
with $\tilde m_q$, an average value of squark masses.\footnote{
The CP violation parameter $\epsilon_K$ gives about one order 
severer
constraints on the imaginary part of $(\delta_{12}^D)_{XY}$ than
the real part.
We here concentrate ourselves only on the constraints from the real
part of $K^0\bar K^0$ mixing, 
since under the 
other experimental constraints to the CP phase originated from 
SUSY breaking sector, which are mainly given by electric dipole moment,
we may expect that the CP phases are small enough to satisfy the 
constraints from the imaginary part of the $K^0\bar K^0$ mixing.
}
The $\mu\rightarrow e \gamma$ process gives
\begin{equation}
\ |(\delta_{12}^E)_{LL}|,|(\delta_{12}^E)_{RR}|\leq 3.8\times 10^{-3}
\left( \frac{\tilde m_l ({\rm GeV})}{100}\right)^2,\label{LLl}
\label{LFV}
\end{equation}
where $\tilde m_l$ is an average mass of scalar leptons.
In the usual anomalous $U(1)_A$ scenario, $\Delta$ can be
estimated as
\begin{equation}
(\Delta_{ij}^F)_{XX}\sim
\lambda^{|f_i-f_j|}(|f_i-f_j|)\VEV{D_A},
\end{equation}
since the mass difference is given by $(f_i-f_j)\VEV{D_A}$,
where $f_i$ is the anomalous $U(1)_A$ charge of $F_i$.
Here the reason for appearing the coefficient $\lambda^{|f_i-f_j|}$ 
is that the unitary diagonalizing
matrices are given by
\begin{equation}
\left(
\begin{array}{cc}
1 & \lambda^{|f_i-f_j|} \\
-\lambda^{|f_i-f_j|} & 1
\end{array}
\right).
\end{equation}
Therefore if the condition $\psi_1=t$ is satisfied,
%In our scenario, the anomalous $U(1)_A$ charge of ${\bf \bar 5}_1$
%is the same as that of ${\bf \bar 5}_2$, 
namely the sfermion masses of ${\bf \bar 5}_1$ and 
${\bf \bar 5}_2$ are almost degenerate, 
the constraints from these FCNC processes become weaker. 
This is because the constraints from the $K^0-\bar K^0$ 
mixing and the CP violation to the product 
$(\delta_{12})_{LL}\times (\delta_{12})_{RR}$ are much stronger
than those to $(\delta_{12})_{LL}^2$ or $(\delta_{12})_{RR}^2$
as shown in eq. (\ref{LR}) and (\ref{LL}).
Therefore suppression of $(\Delta_{12}^D)_{RR}$ makes the constraints
much weaker. 

In the next section, we show that in $E_6$ unification, the above condition
is automatically satisfied.

\section{$E_6$ unification}
%Let us first  recall the twisting mechanism,
%which has been proposed by one of the authors (M.B.).
%\cite{Bando}
%The twisting family structure by this mechanism is peculiar to
%$E_6$ unification model and we here explain how it happens.
In the case of $E_6$, ${\bf 16}$ and ${\bf 10}$ of $SO(10)$ are naturally
included in a single multiplet {\bf 27} of $E_6$.
The fundamental representation of $E_6$ contains {\bf16} and
{\bf10} of $SO(10)$ automatically:
Under $E_6\supset SO(10)\supset SU(5)$,
\begin{equation}
{\bf27} \rightarrow \underbrace{[( {\bf 16,10}) +({\bf 16,\bar 5})
+({\bf 16,1})]}_{\bf 16}
+\underbrace{[({\bf 10,\bar 5})+({\bf 10,5})]}_{\bf 10}
+ \underbrace{[({\bf 1,1})]}_{{\bf 1}}
\end{equation}
where the representation of $SO(10), SU(5)$ are explicitly denoted
in the above.
Therefore the $E_6$ model naturally has the freedom
for replacing matter fields $({\bf 16,\overline 5})$ by
$({\bf 10,\overline 5})$.
%So here let us explain how the twisting family structure arises
%in the $E_6$ unification.
In order to see how the replacement happens, we introduce the following
Higgs fields
%\footnote{
%Note that
%additional Higgs fields $\bar \Phi ({\bf \overline{27}})$ and
%$\bar C({\bf \overline{27}})$ are required for satisfying
%$D$-flatness condition of $E_6$ gauge theory, and an adjoint
%field $A({\bf 78})$ to break the GUT gauge group into the 
%standard gauge group. In order to realize doublet-triplet
%splitting, the actual breaking pattern must be
%$E_6\rightarrow SO(10) \rightarrow 
%SU(3)_C\times SU(2)_L\times SU(2)_R\times U(1)_{B-L} 
%\rightarrow SU(3)_C\times SU(2)_L \times U(1)_Y$.\cite{maekawa,BKM}
%} 
which are relevant to determine the mass matrices
of matter multiplets $\Psi_i({\bf 27})$, whose  $U(1)_A$ charges 
are denoted as  $\psi_i$\footnote{
We assume that $\psi_1>\psi_2>\psi_3$}(i=1,2,3):
\begin{enumerate}
\item $\Phi({\bf 27})$ and $\bar \Phi({\bf \overline{27}})$:
$\VEV{\Phi}=\VEV{\bar \Phi}
=\lambda^{-(\phi+\bar \phi)}$ break $E_6$ into $SO(10)$,
\item $C({\bf 27})$ and $\bar C({\bf \overline{27}})$:
$\VEV{C}=\VEV{\bar C}=\lambda^{-(c+\bar c)}$
break $SO(10)$ into $SU(5)$,
\item  $H({\bf 27})$:Higgs field  which includes the Higgs doublets.
\end{enumerate}
The $U(1)_A$ invariant superpotential for low energy Yukawa terms is, 
\begin{equation}
W_Y=\left(\frac{\Theta}{\Mp}\right)^{\psi_i+\psi_j+h}\Psi_i\Psi_jH,
\end{equation}
and that for the replacement is
\begin{equation}
W=\lambda^{\psi_i+\psi_j+\phi}\Psi_i\Psi_j\Phi
+\lambda^{\psi_i+\psi_j+c}\Psi_i\Psi_j C,
\end{equation}
where we suppress the coefficients of order one and
for the above we assume that $\psi_i+\psi_j+h\geq 0$
for each $i,j$ pair so that there appears no SUSY zero.

The VEVs $\VEV{\Phi}=\VEV{\bar \Phi}=\lambda^{-(\phi+\bar \phi)}$ 
and $\VEV{C}=\VEV{\bar C}=\lambda^{-(c+\bar c)}$ induce
the masses between ${\bf 5}$ and ${\bf \bar 5}$ as
\begin{equation}
\bordermatrix{
&\Psi_1({\bf 16,\bar 5})&\Psi_2({\bf 16,\bar 5})
&\Psi_3({\bf 16,\bar5})
&\Psi_1({\bf 10, \bar 5})&\Psi_2({\bf 10, \bar5})
&\Psi_3({\bf 10,\bar 5})\cr
\Psi_1({\bf 10,5})&\lambda^{2\psi_1+r}&\lambda^{\psi_1+\psi_2+r}
&\lambda^{\psi_1+\psi_3+r}
&\lambda^{2\psi_1}&\lambda^{\psi_1+\psi_2}
&\lambda^{\psi_1+\psi_3}  \cr
\Psi_2({\bf 10,5})&\lambda^{\psi_1+\psi_2+r}&\lambda^{2\psi_2+r}
&\lambda^{\psi_2+\psi_3+r}
& \lambda^{\psi_1+\psi_2}   &  \lambda^{2\psi_2}
&\lambda^{\psi_2+\psi_3} \cr
\Psi_3({\bf 10,5}) &\lambda^{\psi_1+\psi_3+r}&\lambda^{\psi_2+\psi_3+r}
&\lambda^{2\psi_3+r}
 &\lambda^{\psi_1+\psi_3} & \lambda^{\psi_2+\psi_3}
 & \lambda^{2\psi_3}  \cr} \lambda^{\frac{1}{2}(\phi-\bar \phi)},
 \label{full}
\end{equation}
where we define  a parameter $r$ as 
\begin{equation}
\lambda^r\equiv \lambda^{\frac{1}{2}(c-\bar c-\phi+\bar \phi)}.
\end{equation}
Since $\psi_3<\psi_1,\psi_2$, $\Psi_3$ has larger masses than 
$\Psi_1$ and $\Psi_2$. Therefore 3 massless modes tends to consist of
$\Psi_1$ and $\Psi_2$. Actually under some conditions, 
the 3 massless modes become
\begin{eqnarray}
{\bf \overline 5}_1 &=& \Psi_1({\bf 16},{\bf \overline 5})
+\lambda^{\psi_1-\psi_3}\Psi_3({\bf 16},{\bf \overline 5})
+\lambda^{\psi_1-\psi_2+r}\Psi_2({\bf 10},{\bf \overline 5})
+\lambda^{\psi_1-\psi_3+r}\Psi_3({\bf 10},{\bf \overline 5}), 
\label{51} \\
{\bf \overline 5}_2 &=& \Psi_1({\bf 10},{\bf \overline 5})
+\lambda^{\psi_1-\psi_3-r}\Psi_3({\bf 16},{\bf \overline 5})
+\lambda^{\psi_1-\psi_2}\Psi_2({\bf 10},{\bf \overline 5})
+\lambda^{\psi_1-\psi_3}\Psi_3({\bf 10},{\bf \overline 5}), 
\label{52} \\
{\bf \overline 5}_3 &=& \Psi_2({\bf 16},{\bf \overline 5})
+\lambda^{\psi_2-\psi_3}\Psi_3({\bf 16},{\bf \overline 5})
+\lambda^{r}\Psi_2({\bf 10},{\bf \overline 5})
+\lambda^{\psi_2-\psi_3+r}\Psi_3({\bf 10},{\bf \overline 5}),
\label{53}
\end{eqnarray}
where the first terms of the right hand side are the main components of
these massless modes and the other terms are mixing terms with
heavy states, $\Psi_3({\bf 16},{\bf \overline 5})$,
$\Psi_2({\bf 10},{\bf \overline 5})$ and
$\Psi_3({\bf 10},{\bf \overline 5})$.
This is almost the same situation as discussed
in the previous section. Actually if we take $r=1/2$, namely,
\begin{equation}
1=c-\bar c-\phi+\bar \phi,
\end{equation}
the massless modes discussed in the previous section are obtained.
Namely, all the quark and lepton mass matrices are obtained even in 
this $E_6$ unification. 
Only the difference is that in $E_6$ unification the charge of the main
part of second generation ${\bf \bar 5_2}$ is fixed as $\psi_1$. 
Therefore the condition for suppression of $K^0\bar K^0$ mixing, which
was discussed in the previous section, is automatically satisfied.

Now that the constraints from the $K^0\bar K^0$ mixing (and the 
CP violation) become weaker as discussed above, we have larger region
in the  
paramter space, where the lepton flavor
violating processes like $\mu\rightarrow e \gamma$ are appreciable.
Actually, if the ratio of the VEV of $D_A$ to the gaugino mass squared 
at the GUT scale is given by
\begin{equation}
R\equiv \frac{\VEV{D_A}}{M_{1/2}^2},
\end{equation}
the scalar fermion mass square at the low energy scale is estimated as
\begin{equation}
\tilde m_{F_i}^2\sim f_i R M_{1/2}^2+\eta_FM_{1/2}^2,
\end{equation}
where $\eta_F$ is a renormalization group factor.
Therefore in our scenario, the eq.(\ref{LL}) for 
$(\delta_{12}^D)_{LL}$
becomes
\begin{eqnarray}
(\delta_{12}^D)_{LL}&\sim &
\lambda\frac{(\psi_1-\psi_2)R M_{1/2}^2}
{(\eta_{D_L}+\frac{\psi_1+\psi_2}{2}R)M_{1/2}^2}
=\lambda\frac{(\psi_1-\psi_2)R}{(\eta_{D_L}+\frac{\psi_1+\psi_2}{2}R)} \\
&\leq &4.0\times 10^{-2}
\left(\frac{(\eta_{D_L}+\frac{\psi_1+\psi_2}{2}R)^{1/2}M_{1/2}({\rm GeV})}
           {500}\right),
\end{eqnarray}
which is rewritten
\begin{equation}
M_{1/2}\geq 
1.25\times 10^4 \lambda\frac{(\psi_1-\psi_2)R}
{(\eta_{D_L}+\frac{\psi_1+\psi_2}{2}R)^{3/2}}
({\rm GeV}).
\end{equation}
Though the main contribution to $(\delta_{12}^D)_{RR}$ vanishes, through
the mixing in eq. (\ref{51}) and (\ref{52}), $(\delta_{12}^D)_{RR}$ is
estimated as
\begin{equation}
(\delta_{12}^D)_{RR}\sim \lambda^{\frac{1}{2}}\frac{\lambda^2(-\psi_2)R}
                                                  {\eta_{D_R}+\psi_1R},
\end{equation}
where the mixing $\lambda^{\frac{1}{2}}$ is different from the naively
expected value $1=\lambda^{\psi_1-\psi_1}$.
From the eq.(\ref{LR}) for $\sqrt{(\delta_{12}^D)_{LL}(\delta_{12}^D)_{RR}}$,
the constraint to the gaugino mass $M_{1/2}$ is given by
\begin{equation}
M_{1/2}\geq 1.8\times 10^5\frac{\lambda^{1.75} R\sqrt{\psi_2(\psi_1-\psi_2)}}
                               {(\eta_D +\psi_1 R)^{1.5}}.
\end{equation}
%Similarly, the constraint of Im($(\delta_{12}^D)_{LL}$)
%gives 12.5 times severer limit than that of the real part.
On the other hand, the eq.(\ref{LFV}) for $(\delta_{12}^E)_{RR}$ leads to
\begin{equation}
M_{1/2}\geq 1.6\times 10^3
\frac{(\lambda (\psi_1-\psi_2)R)^{1/2}}
{\eta_{E_R}+\frac{\psi_1+\psi_2}{2}R}
({\rm GeV}).
\end{equation}
Taking probable values, $\psi_1=5$, $\psi_2=4$, 
$\eta_{D_L}\sim \eta_{D_R}\sim 6$ and 
$\eta_{E_R}\sim 0.15$, 
the lower limits of the gaugino mass are roughly estimated as in
Table.1.
\begin{center}
\begin{tabular}{|c|c|c|c|c|c|} 
\hline
$R$                              & 0.1  & 0.3 & 0.5 & 1 & 2 \\ \hline
$(\delta_{12}^D)_{LL}$   & 15 & 38  & 53  & 73 & 86 \\ \hline
$\sqrt{(\delta_{12}^D)_{LL}(\delta_{12}^D)_{RR}}$   
                         & 120 & 300 & 420 & 560 & 690 \\ \hline
$|(\delta_{12}^E)_{RR}| $          & 370 & 260 & 210 & 150 & 110\\ \hline
\end{tabular}

\vspace{5mm}
Table 1. Lower bound of gaugino mass $M_{1/2}$ at GUT scale (GeV).
\end{center}
Note that when $R=0.1$, the $\mu\rightarrow e\gamma$
process gives the severest constraint in these FCNC 
processes\cite{kurosawa}.
Therefore the lepton flavor violating processes\cite{kurosawa,LFV} 
might be seen in future, though the prediction is strongly dependent on
the detail of the SUSY breaking sector.

The reason for suppression of $(\Delta_{12}^D)_{RR}$
is that the anomalous $U(1)_A$ charge of ${\bf \overline 5}_2$ becomes the
same as that of ${\bf \overline 5}_1$ because the fields ${\bf \overline
5}_1$
and ${\bf \overline 5}_2$ are originated from a single field $\Psi_1$.
This is a non-trivial situation. The massless mode of the second 
generation
${\bf \overline 5}_2=\Psi_1({\bf 10},{\bf \bar 5})
+\lambda^{5/2}\Psi_3({\bf 16},{\bf \bar 5})$
has Yukawa couplings through the second term
$\lambda^{5/2}\Psi_3({\bf 16},{\bf \bar 5})$. However, for SUSY breaking
term
which is proportional to the anomalous $U(1)_A$ charge, the contribution
from the first term dominates the one from the second term, 
which realizes the
degenerate SUSY breaking terms between the first and the second generation.
It is suggestive that the requirement to reproduce the bi-large mixing
angle in neutrino sector leads to this non-trivial structure, which
suppresses the FCNC processes.\footnote{
We should comment on $D$-term contribution to the scalar fermion
masses. Generically such $D$-term has non-vanishing VEV
\cite{kawamura} when the rank of
the gauge group is reduced by the symmetry breaking and SUSY breaking terms
are non-universal. In our scenario,
when $E_6$ gauge group is broken to $SO(10)$ gauge group, the $D$-term
contribution gives different values to the sfermion masses of {\bf 16} 
and {\bf 10} of $SO(10)$, which destroys the natural suppression of FCNC
in the $E_6$ unification. However, if SUSY breaking parameters become
universal by some reason, the VEV of $D$ can become negligible. 
Actually, the condition $m_\phi^2=m_{\bar \phi}^2$ makes the VEV of
the $D$ much suppressed.
Therefore in principle, we can control the $D$-term contribution, though
it is dependent on the SUSY breaking mechanism. 
}
In this way, such a non-trivial structure is automatically
obtained in $E_6$ model, which is much different from 
the $SO(10)$ model in which the condition
can be satisfied only by hand.

\section{Conclusion}
In this talk, we proposed a GUT scenario of $SO(10)$ unified model 
in which
 DT splitting is naturally realized by the DW mechanism. 
The anomalous $U(1)_A$ gauge symmetry plays an essential role in the DT 
splitting.
Using this mechanism, we examined the simplest model in which realistic
mass matrices of quarks and leptons,
including the neutrino, can be determined by the anomalous $U(1)_A$ charges.
This model predicts bi-maximal mixing angles in the neutrino sector, a small 
value of $\tan \beta$, and the relation $V_{e3}\sim \lambda$. 
Proton stability is naturally 
realized.
It is interesting that once we fix
the anomalous $U(1)_A$ charges for all fields, 
the order of each parameter and 
scale is determined, except that of the SUSY breaking.

Extension into $E_6$ unification\cite{BM} is also interesting 
that the mass matrices with bi-maximal mixing discussed in 
this paper appear again in the $E_6$ unified model. Moreover, the
condition $\psi_1=t$, which makes the constraints from the FCNC process 
weaker,
is automatically satisfied.
In subsequent paper\cite{BM2}, it is shown that the DT splitting mechanism 
can be non-trivially incorporated into $E_6$ unification. 

It is very suggestive that the anomalous $U(1)_A$ gauge symmetry 
motivated
by superstring theory plays a critical role in solving the two 
biggest 
problems in GUT, the fermion mass hierarchy problem and the 
doublet-triplet
splitting problem. This may be the first evidence for the 
validity of string theory from the phenomenological point of view.

\section{Acknowledgements}
We would like to thank M. Bando and T. Kugo for interesting discussions
and useful comments.

\end{document}